\documentclass[twocolumn,showpacs,preprintnumbers,amsmath,amssymb,letterpaper]{revtex4}

\usepackage{graphicx} 
\usepackage{dcolumn}  
\usepackage{bm}       

\begin{document}

\title{Dilute Bose gases interacting via power-law potentials}

\author{Ryan M. Kalas and D. Blume}
\affiliation{
Department of Physics and Astronomy, Washington State University, Pullman,
Washington 99164-2814}


\begin{abstract}
Neutral atoms interact through a van der Waals potential which
asymptotically falls off as $r^{-6}$. 
In ultracold gases, this interaction
can be described to a good approximation by the atom-atom
scattering length.  However, corrections arise that depend
on the characteristic length of the van der Waals potential.  
We parameterize these corrections by analyzing the energies of 
two- and few-atom systems under external harmonic confinement,
obtained by numerically and analytically solving the 
Schr\"odinger equation. 
We generalize our results to particles interacting through 
a longer-ranged potential which asymptotically falls off as $r^{-4}$.

\end{abstract}

\maketitle


\section{Introduction}

The interaction strengths of sufficiently dilute and cold bosonic
atom samples such as Bose-Einstein condensates of alkali atoms can
be parameterized to a good approximation by a single parameter,
the $s$-wave scattering length~\cite{pethickbook}. In these systems, the
neutral atoms interact through short-ranged van der Waals
potentials which fall off as $r^{-6}$ at large interparticle
distances $r$. More recently, progress has been made in cooling
and trapping systems characterized by interaction potentials that
fall off more slowly than $r^{-6}$ at large $r$. For example, the
interaction between a neutral atom and an ion is dominated by a
polarization potential that falls off asymptotically as 
$r^{-4}$~\cite{bransden}.
Atom-ion systems have recently been proposed as candidates for
quantum computing applications~\cite{zbig2007}, 
and also play a role in recent
work which proposes that macroscopic molecules can be formed by
immersing an ion in a condensed Bose gas~\cite{cote,massignan,collin}. 
Another example for
systems with longer-ranged interactions are dipolar
gases~\cite{santos,pfau}. 
In these systems, the non-neglegible magnetic or electric
dipole moment leads to an angle-dependent $r^{-3}$ potential at
large interparticle distances.
A natural question to ask is how
well the properties of Bose systems with longer-ranged
interactions can be described by the $s$-wave scattering length.

This paper considers dilute bosonic systems under external
confinement interacting through spherically symmetric power-law
potentials. In particular, we treat interactions with
$r^{-n}$ tails, where $n$ is $4$ or $6$. We focus on the regime
where the characteristic length $\beta_n$ of the two-body
potential is much smaller than the characteristic length $a_{ho}$
of the trapping potential. In this regime, the
shape-dependent interaction potential can be replaced by a
regularized zero-range potential whose interaction strength is
parametrized by the $s$-wave scattering length. For the
potential with $r^{-6}$ tail, e.g., it has been shown previously
that the energy levels of the trapped two-body system can be
reproduced very acccurately if the energy-dependence of the scattering
length is accounted for \cite{blume,bolda}. This paper 
extends the two-body analysis to
potentials with $r^{-4}$ tail, whose scattering length
has---because of the longer-ranged character of the potential---a
stronger energy-dependence than that of potentials with
$r^{-6}$ tail.  We find that the corrections 
to the energy predicted by the zero-energy scattering length 
go as $(\beta_6 / a_{ho})^3$ and $(\beta_4 / a_{ho})^2$ for the
interaction potentials with $r^{-6}$ and $r^{-4}$ tails, respectively.

Using Monte Carlo techniques, we furthermore treat
dilute bosonic many-body systems. As in the two-body case, we
consider different interaction potentials and analyze the
resulting eigenenergies. Not unexpectedly, our results show that
the energy-dependent scattering length remains a good quantity
also in the many-body system. This suggests, e.g., that the
description of dilute Bose gases within a mean-field
Gross-Pitaevskii framework can be improved notably by including
the energy-dependence of the scattering length. First steps in 
this direction have already been taken~\cite{gao03,collin}; 
our results provide 
additional benchmark results that may aid in further assessing 
the accuracy of these and related frameworks.

Section~\ref{sectionII} introduces the Hamiltonian and the model interaction
potentials used in our study. Section~III discusses the energetics
of two particles in a trap interacting through both finite-range
and zero-range potentials. In Sec.~IV, we consider the
energetics of more than two particles in a trap by solving the
many-body Schr\"odinger equation using Monte Carlo techniques.
Finally, Sec.~V concludes.

\section{Hamiltonian}
\label{sectionII}

The Hamiltonian for a system consisting of $N$ identical mass $m$
bosons in the presence of a spherically symmetric harmonic
trapping potential with angular frequency $\omega$ is given
by
\begin{equation}
\label{hb}
H =  \sum_{i=1}^N\Big{(}-\frac{\hbar^2}{2m}\nabla^2_i+\frac{1}{2}m\omega^2
\mathbf{r}_i^2\Big{)}+\sum_{i<j}^N v(r_{ij}),
\end{equation}
where $\mathbf{r}_i$ denotes the position vector of the {\it i}th
atom. The spherically symmetric two-body interaction potential $v$
depends on the relative distance $r_{ij}$, $r_{ij}=|\mathbf{r}_i -
\mathbf{r}_j|$.  We consider attractive power-law potentials
with a hardcore radius $r_c$,
\begin{equation}
\label{vn}
v_{n}(r) = \left\{
\begin{array}{c l}
  \infty       &  \mbox{ for } \quad  r<r_c \,   \\
  -C_n/r^n     &  \mbox{ for } \quad  r>r_c \, ,
\end{array}
\right.
\end{equation}
with $n=4$, $6$ and $C_n>0$.  The Hamiltonian defined in
Eq.~(\ref{hb}) is characterized by three length scales: the
hardcore radius $r_c$, the van der Waals length scale $\beta_n$
[$\beta_n= (m C_n/\hbar^2)^{1/(n-2)}$], and the harmonic oscillator
length $a_{ho}$ [$a_{ho}=\sqrt{\hbar/m\omega}$].  Throughout
this paper, we are interested in the regime where $r_c$ and
$\beta_n$ are much smaller than $a_{ho}$.

In three dimensions, the interaction strength of any potential
that falls off faster than $r^{-3}$ at large distances can be
characterized by the energy-dependent free-space $s$-wave
scattering length $a(k)$~\cite{newton},
\begin{equation}
\label{ascatt}
a(k)=-\frac{\tan \delta(k)}{k},
\end{equation}
where $\delta(k)$ denotes the $s$-wave scattering phase shift and
$k$ the wave vector at the scattering energy of $\hbar^2 k^2/m$.  
The zero-energy scattering length $a(0)$ is 
defined by taking the $k\rightarrow0$ limit of
Eq.~(\ref{ascatt}). For the $v_4$ and $v_6$ potentials, the
zero-energy and energy-dependent scattering lengths can be
calculated from analytical solutions derived using series
expansion techniques \cite{holzwarth,gaor6}.

Figures \ref{cap1}(a) and \ref{cap2}(a) show the zero-energy
scattering length $a(0)$ as a function of $\beta_n$ for the
$v_n$ potential with $r_c=0.007a_{ho}$
for $n$ equals $4$ and $6$, respectively.  Although the scattering
lengths are calculated for the free-space system with no external
trapping potential, we choose to express all lengths in
units of $a_{ho}$ to ease the comparison with the trapped system
in Secs.~III and IV.  When $\beta_n=0$, the scattering length
coincides with the hardcore radius $r_c$. As $\beta_n$ increases,
the attractive tails of the $v_n$ potentials increase in
strengths, which leads to a decrease of the scattering lengths.
This continues until the potential is strong enough to support its
first bound state, at which point the scattering length changes
its sign from negative to positive.  Figures \ref{cap1}(a) and
\ref{cap2}(a) indicate that this divergence occurs at different
values of $\beta_n$, i.e., at $\beta_4\approx0.022 a_{ho}$ and
$\beta_6\approx0.016 a_{ho}$, owing to the fact that the $v_6$
potential is shorter-ranged than the $v_4$ potential.  This can be
understood heuristically by considering the ratio of the
attractive power of the potentials, that is, the ratio $\int
v_6(r) {\rm d}^3\mathbf{r}/ \int v_4(r) {\rm d}^3\mathbf{r}$.
Taking the limits of 
the $r$-integration 
as $r_c$ and $\infty$, this ratio
equals $\beta_6^4/(3\beta_4^2r_c^2)$. Looking at equal values of
$\beta_4$ and $\beta_6$ and considering that $\beta_n>r_c$ at the
first divergence, this ratio is greater than one, in agreement
with the observation that the $v_6$ potential supports an $s$-wave
bound state for smaller values of $\beta_n/a_{ho}$ than the $v_4$
potential.

Throughout this paper we are interested in describing dilute Bose
systems which interact primarily through binary $s$-wave collisions.
In such systems, the short-range details of the interaction potential
are not being probed, and the regularized zero-range pseudopotential
$v_{ps}(r)$~\cite{huang},
\begin{equation}
\label{vpseudo}
v_{ps}(r)=\frac{4\pi\hbar^2}{m}g\delta^{(3)}(\mathbf{r})\frac{\partial}{\partial r}r,
\end{equation}
reproduces many observables obtained for the true shape-dependent interaction
potential---in our case, the $v_4$ or $v_6$ potential---accurately if
the strength $g$ is chosen properly.  In Secs.~III and IV we take $g$ to be
the zero-energy scattering length $a{(0)}$ and the energy-dependent
scattering length $a{(k)}$ of the shape-dependent interaction potential $v_n$.

For two particles in a harmonic trap interacting through $v_{ps}$, the Schr\"odinger
equation can be solved analytically~\cite{buschenglert}.  
The center of mass energy equals 
$(n_{cm}+3/2)\hbar\omega$ ($n_{cm}=0$, $1$,$\ldots$), and the 
$s$-wave
eigenenergies
$E_{rel}=\varepsilon\hbar\omega$ of the Schr\"odinger 
equation in the relative coordinate are determined by \cite{buschenglert}
\begin{equation}
\label{busch}
\frac{g}{a_{ho}}=\frac{\Gamma(-\frac{\varepsilon}{2}+\frac{1}{4})}
   {\sqrt{2} \, \Gamma(-\frac{\varepsilon}{2}+\frac{3}{4})}.
\end{equation}
The transcendental equation (\ref{busch}) can be solved straightforwardly 
for any given $g$ using standard root-finding procedures.  For $N>2$, analytical
solutions to the Schr\"odinger equation for trapped atoms interacting
through $v_{ps}$ are in general not known, and we instead resort to numerical techniques
(see Sec.~IV).

\section{Two particles in a trap}
\label{sectionIII}

We first consider the Hamiltonian given in Eq.~(\ref{hb}) with $v=v_n$ 
for $N=2$.
\begin{figure}[hbtp]
\includegraphics[scale=0.2]{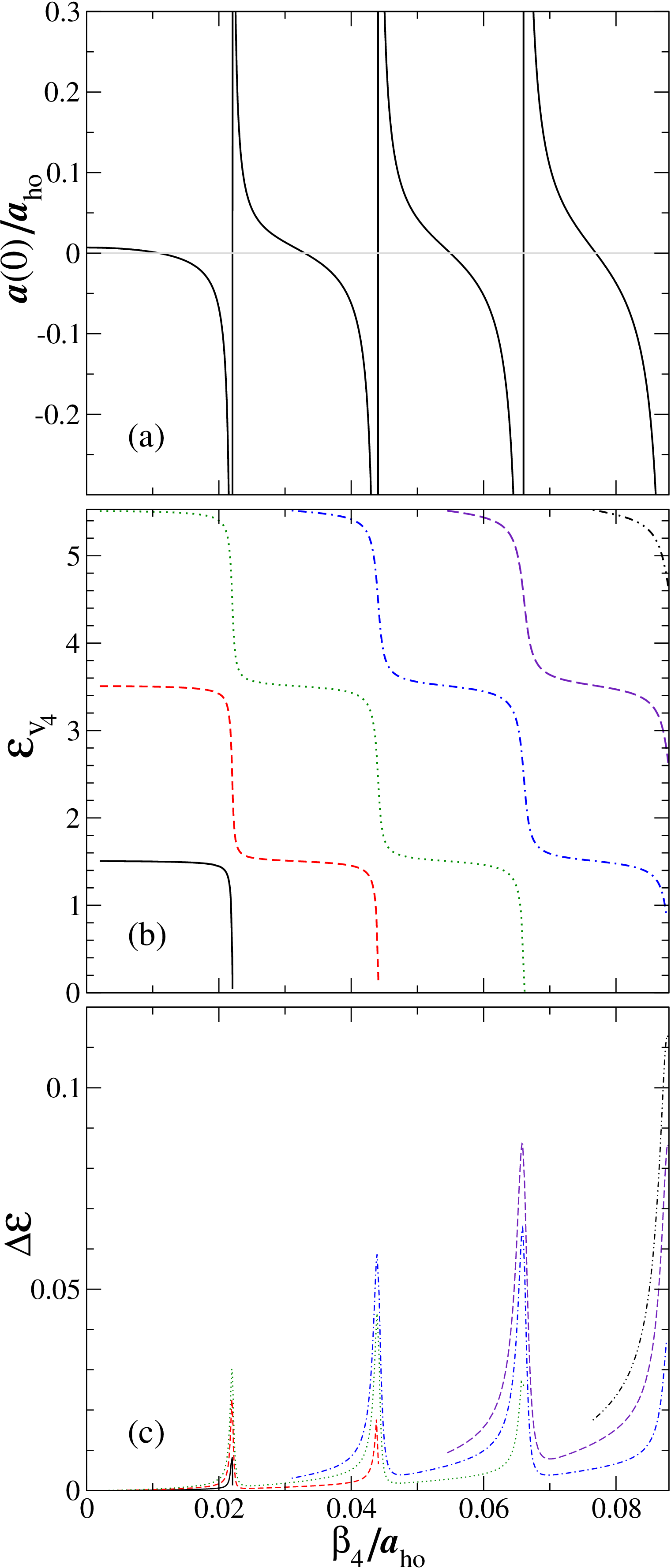}\\  
\caption{\label{cap1}
(Color online)
$s$-wave 
properties of two particles interacting
through the potential $v_4$ with 
$r_c=0.007a_{ho}$ as a function of $\beta_4/a_{ho}$ (note that the
$x$-axis is the same for all three panels): (a) Free-space
zero-energy scattering length $a(0)$. (b) Relative energy
$\varepsilon_{v_4}$ for two trapped atoms. (c)
Energy difference $\Delta \varepsilon$, $\Delta \varepsilon
=\varepsilon_{v_4}-\varepsilon_{a(0)}$, for two
trapped atoms. In (b) and (c), the line styles are keyed to each
other for ease of comparison.
}
\end{figure}
After separating off the center of mass motion, we are left with a Schr\"odinger
equation in the relative coordinate.  We solve the corresponding 
one-dimensional differential equation numerically using $B$-splines.
Figures \ref{cap1}(b) and \ref{cap2}(b) show the resulting 
relative $s$-wave
eigenenergies, denoted by $\varepsilon_{v_4}$ and $\varepsilon_{v_6}$, 
respectively, as a function of $\beta_n$.  As in Figs.~\ref{cap1}(a)
and \ref{cap2}(a), the hardcore radius is fixed at $r_c=0.007a_{ho}$.
For those $\beta_n$ values for which $a(0)$ is small
[see Figs.~\ref{cap1}(a) and \ref{cap2}(a)], the eigenenergies
$\varepsilon_{v_n}$ coincide approximately with the eigenenergies
$(2n_{rel}+3/2)$ 
of the non-interacting system, 
where $n_{rel}=0$, $1$,$\ldots$.
However, each time $a(0)$ diverges, a new molecular state
appears and the energy of the gas-like state decreases by 
approximately $2\hbar\omega$.

Next, we consider two trapped particles interacting through $v_{ps}$.
We find that the zero-range pseudopotential with energy-dependent scattering
length reproduces the eigenenergies $\varepsilon_{v_n}$ for the shape-dependent
potential $v_n$ with high accuracy for all interaction strengths considered
in Figs.~\ref{cap1} and \ref{cap2}.  To obtain the eigenenergies for
$v_{ps}$ with $g=a(k)$, which we denote by $\varepsilon_{a(\varepsilon)}$,
we calculate $a(k)$ for different $\beta_n$ and solve Eq.~(\ref{busch})
self-consistently \cite{blume,bolda}, i.e., 
we require that 
$\varepsilon \hbar \omega$ 
on the right-hand side
of Eq.~(\ref{busch}) agrees with the energy $\hbar^2 k^2 /m$ at which the two
particles collide.  
Since $\varepsilon_{v_n}$ and $\varepsilon_{a(\varepsilon)}$
coincide to many digits, Eq.~(\ref{busch}) can be used to describe the physics
of two trapped particles provided $\beta_n \ll a_{ho}$ and provided the
energy-dependence of the scattering length is known.  
For short-range potentials, this was 
already shown in 
Refs.~\cite{blume,bolda}.
For some systems under experimental study, only $a(0)$ is known 
[$a(k)$ is unknown].  It is thus useful to quantify the deviations 
$\Delta \varepsilon$ between the eigenenergies $\varepsilon_{v_n}$ and 
the eigenenergies $\varepsilon_{a(0)}$ obtained from Eq.~(\ref{busch})
with $g=a(0)$.

\begin{figure}[hbtp]
\includegraphics[scale=0.2]{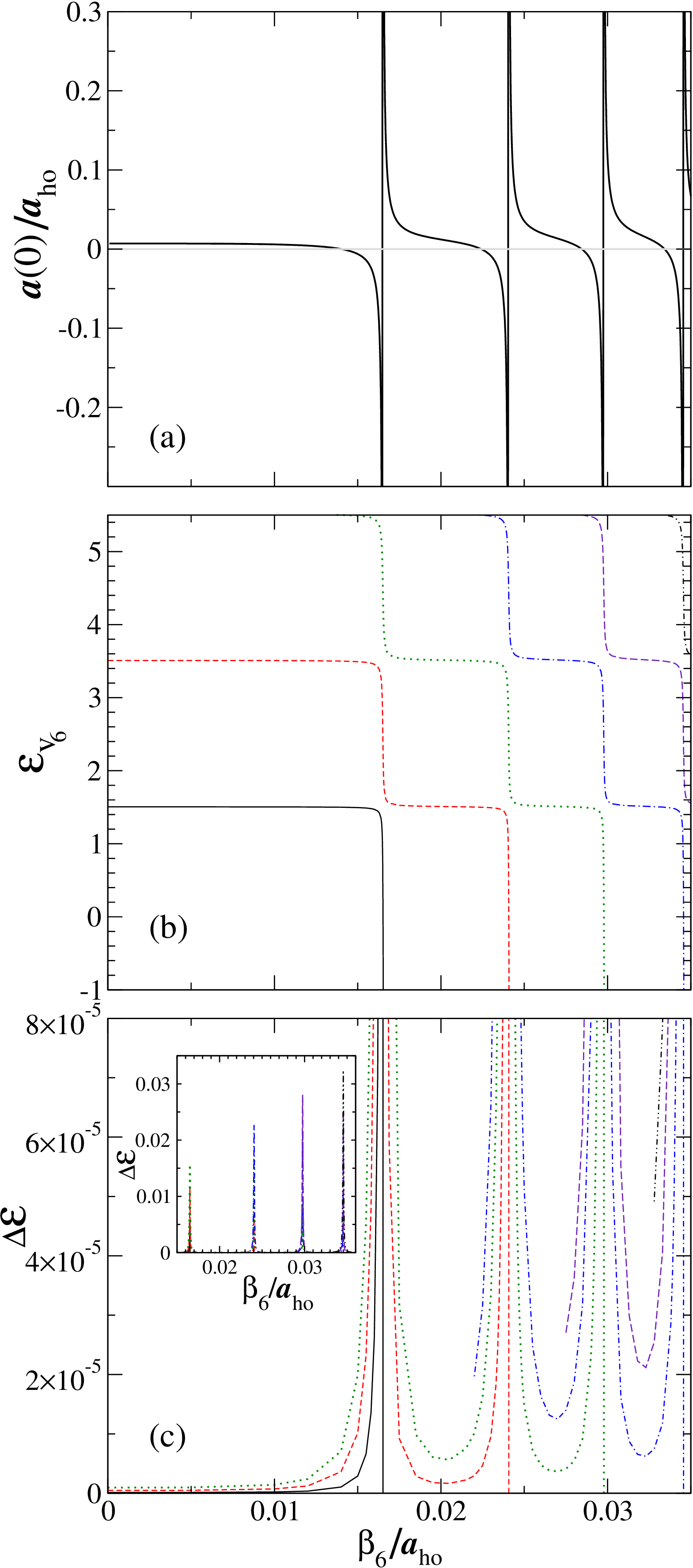}\\  
\caption{\label{cap2} 
(Color online)
$s$-wave 
properties of two particles interacting
through the potential $v_6$ with 
$r_c=0.007a_{ho}$ as a function of $\beta_6/a_{ho}$ (note that the
$x$-axis is the same for all three panels): (a) Free-space
zero-energy scattering length $a(0)$. (b) Relative energy
$\varepsilon_{v_6}$ for two trapped atoms. (c) Energy difference 
$\Delta \varepsilon$, $\Delta \varepsilon=\varepsilon_{v_6}-\varepsilon_{a(0)}$, 
for two trapped atoms. The inset of (c) plots $\Delta \varepsilon$ 
on an enlarged scale to show the maximum of $\Delta \varepsilon$.
In (b) and (c), the lines styles are keyed to each other for ease
of comparison. }
\end{figure}
Figures~\ref{cap1}(c) and \ref{cap2}(c)
show the energy difference $\Delta \varepsilon$, 
$\Delta \varepsilon = \varepsilon_{v_n} - \varepsilon_{a(0)}$,
for the three energetically lowest-lying gas-like states.  The line styles in
Figs.~\ref{cap1}(c) and \ref{cap2}(c) correspond to those used
in Figs.~\ref{cap1}(b) and \ref{cap2}(b).  The energy difference
$\Delta \varepsilon$ is larger for the energetically higher-lying 
gas-like states since the energy-dependence of $a(k)$ for a given
$\beta_n$ increases with increasing $\varepsilon$.
The maximum of $\Delta \varepsilon$ increases with increasing 
$\beta_n/a_{ho}$ [see Fig.~\ref{cap1}(c) and the inset of 
Fig.~\ref{cap2}(c)]
and is of the same order of magnitude for the $v_4$ and $v_6$ potentials for
comparable values of $\beta_n/a_{ho}$.  Furthermore, for those $\beta_n$
values for which the scattering length $a(0)$ is comparatively small, the
energy difference $\Delta \varepsilon$ also increases with increasing 
$\beta_n/a_{ho}$.  The magnitude of these ``background'' energy differences
is much larger for the $v_4$ potential than for the $v_6$ potential
[note the difference in the $y$-scales of Figs. \ref{cap1}(c) and \ref{cap2}(c)]. 
For example, for $\beta_n \approx 0.03 a_{ho}$, the background energy is about
$10^{-3}\hbar\omega$ and $10^{-5}\hbar\omega$ for the lowest-lying
gas-like states of the $v_4$ and $v_6$ potentials, respectively.
We now show that the background energy difference $\Delta \varepsilon$
is proportional to
$(\beta_4/a_{ho})^2$ and $(\beta_6/a_{ho})^3$ 
for the $v_4$ and $v_6$ potentials, respectively, thus explaining the 
much smaller energy difference for the $v_6$ potential than for the $v_4$
potential.

\begin{figure}[hbtp]
\includegraphics[scale=0.2]{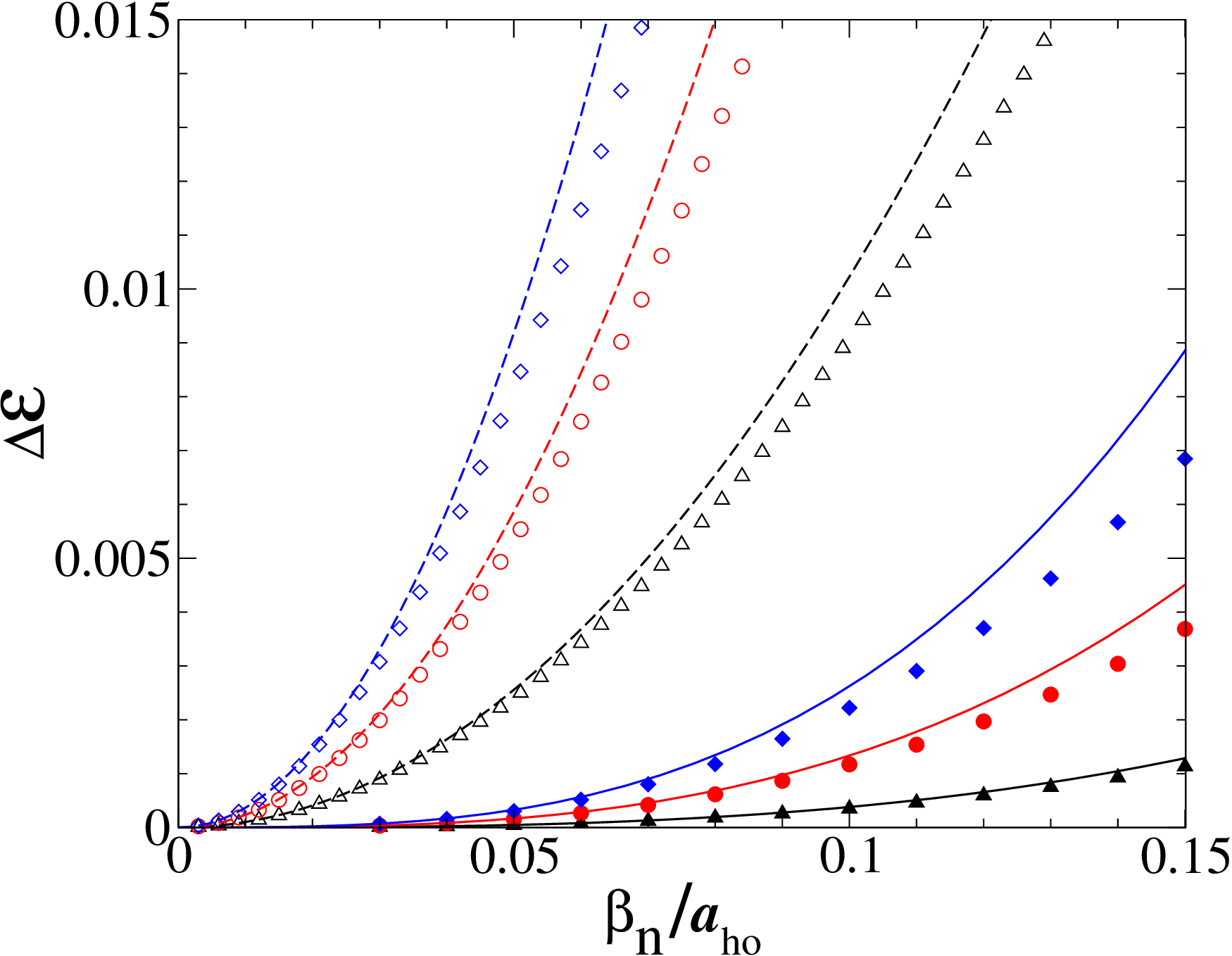}\\
\caption{\label{cap3} 
(Color online)
Energy difference $\Delta\varepsilon$, $\Delta\varepsilon 
= \varepsilon_{v_n} - \varepsilon_{a(0)}$, for the three
energetically lowest-lying gas-like states of two
trapped atoms for $a(0)=0$ as a function of $\beta_n/a_{ho}$.  
The filled (open) triangles, circles
and diamonds show
the numerically determined energy differences for the
levels near $1.5$, $3.5$
and $5.5 \hbar\omega$, respectively,
for the $v_6$ ($v_4$) potential. 
The 
solid and dashed 
lines show the
corresponding analytically determined estimates for
$\Delta\varepsilon$, Eqs. (\ref{sh6x}) and (\ref{sh4x}).
}
\end{figure}

To arrive at these estimates, we use that $\varepsilon_{a(\varepsilon)}$
agrees to many digits with $\varepsilon_{v_n}$,
which implies 
$\Delta\varepsilon = \varepsilon_{a(\varepsilon)}-\varepsilon_{a(0)}$.
Since $\varepsilon_{a(\varepsilon)}$ is determined from Eq.~(\ref{busch})
with $g=a(\varepsilon)$, we can obtain a simple expression for 
$\varepsilon_{a(\varepsilon)}$ by expanding the left-hand side of 
Eq.~(\ref{busch}) about $a(0)$ and the right-hand side about
$\varepsilon_{a(0)}$.  The expansions of $a(k)$ for the $v_4$
and the $v_6$ potential are given by \cite{omalley}
\begin{equation}
\label{aEr4x} a(k)=a(0)+\frac{\pi}{3}\beta_4^2 k +\cdots
\end{equation}
and \cite{newton}
\begin{equation}
\label{aEr6x} a(k) = a(0) \left(1+\frac{1}{2}r_e a(0)
k^2+\cdots \right),
\end{equation}
respectively.  In Eq.~(\ref{aEr6x}), $r_e$ denotes the 
effective range of the $v_6$ potential \cite{gao03,gaoreff},  
\begin{equation}
\label{r6rangex}
\frac{r_{e}}{\beta_6}=\frac{2}{3x_e}\frac{1}{(a(0)/\beta_6)^2}
\Big{[}1+ \left(1-x_e\frac{a(0)}{\beta_6} \right)^2\Big{]},
\end{equation}
where the constant $x_e = [\Gamma(1/4)]^2/(2\pi)\approx2.09$.
Denoting the right-hand side of Eq.~(\ref{busch}) by 
$f(\varepsilon_{a(0)})$ for $g=a(0)$, we find
\begin{equation}
\label{shiftr4x} \Delta \varepsilon \approx \frac{
\frac{\pi}{3}(\frac{\beta_4}{a_{ho}})^2\sqrt{\varepsilon_{a(0)}} }
 { f'(\varepsilon_{a(0)}) - \frac{\pi}{3}(\frac{\beta_4}{a_{ho}})^2
\sqrt{\varepsilon_{a(0)}}
 }
\end{equation}
for the $v_4$ potential and
\begin{equation}
\label{r6deltaE1x} 
\Delta \varepsilon\approx\frac{
\frac{1}{2}\frac{r_{e}}{a_{ho}}(\frac{a(0)}{a_{ho}})^2 \varepsilon_{a(0)} }
 { f'(\varepsilon_{a(0)}) - \frac{1}{2}\frac{r_e}{a_{ho}}(\frac{a(0)}{a_{ho}})^2 
\varepsilon_{a(0)} }
\end{equation}
for the $v_6$ potential.  
The different powers of $\varepsilon_{a(0)}$ in Eqs.~(\ref{shiftr4x}) and (\ref{r6deltaE1x})
follow directly from the linear and quadratic $k$-dependence of the correction
terms in Eqs.~(\ref{aEr4x}) and (\ref{aEr6x}), respectively.

If $a(0)\ll \beta_6$, the square bracket in the expression
for the effective range $r_e$ is approximately equal to 
$2$
and Eq.~(\ref{r6deltaE1x}) reduces to 
\begin{equation}
\label{r6deltaE2x}
\Delta \varepsilon \approx \frac{
\frac{2}{3 x_e} (\frac{\beta_6}{a_{ho}})^3 \varepsilon_{a(0)} }
{ f'(\varepsilon_{a(0)}) - \frac{2}{3 x_e} (\frac{\beta_6}{a_{ho}})^3 \varepsilon_{a(0)} }.
\end{equation}
Furthermore, for small $a(0)$, $\varepsilon_{a(0)}$ is approximately given by
$(3/2+2n_{rel})$ and $f'(\varepsilon_{a(0)})$ is of order $1$ (taking values
of approximately $1.25$, $0.84$, and $0.67$ for $n_{rel}=0$, $1$, and $2$).
The second term in the denominator of Eqs.~(\ref{shiftr4x}) and (\ref{r6deltaE2x})
can thus be dropped provided $\beta_n$ is much smaller than $a_{ho}$.  This yields
\begin{equation}
\label{sh4x}
\Delta \varepsilon \approx \frac{\pi \sqrt{2n_{rel}+3/2}}{3f'(2n_{rel}+3/2)} \Big{(}
\frac{\beta_4}{a_{ho}}\Big{)}^2
\end{equation}
for the $v_4$ potential and
\begin{equation}
\label{sh6x}
\Delta \varepsilon \approx \frac{2 (2n_{rel}+3/2)}{3x_e f'(2n_{rel}+3/2)} \Big{(}
\frac{\beta_6}{a_{ho}}\Big{)}^3
\end{equation}
for the $v_6$ potential.  
Equations~(\ref{sh4x}) and (\ref{sh6x}) can also be derived 
by applying first order perturbation theory to the trapped two-body system
interacting through a zero-range potential (see Sec.~IV).

Solid and dotted lines in Fig.~\ref{cap3}
show the energy differences $\Delta \varepsilon$ predicted by 
Eqs.~(\ref{sh6x})
and (\ref{sh4x}) for $n_{rel}=0$, $1$
and $2$ (from bottom to top) as
a function of $\beta_n/a_{ho}$  
for the $v_6$ and $v_4$ potential, respectively.
For comparison, filled and open symbols
in Fig.~\ref{cap3} show the 
corresponding
numerically determined energy differences
$\Delta \varepsilon$ for the three energetically lowest-lying gas-like
states.  To calculate these energy differences, we fix $r_c$ and $\beta_n$
so that $a(0)=0$, and vary the harmonic oscillator length.
We find that, as long as $r_c$ and $\beta_n \ll a_{ho}$, the results shown in 
Fig.~\ref{cap3} are independent of the number of bound states supported by
the shape-dependent power-law potential $v_n$.  Figure \ref{cap3} illustrates
that the estimates given in Eqs.~(\ref{sh4x}) and (\ref{sh6x}) are quite accurate
for small $a(0)$.  Thus, our derivation shows that the different powers of the characteristic
length scale $\beta_n$, which explain the larger background values of 
$\Delta \varepsilon$ for the $v_4$ potential compared to the $v_6$ potential,
can be traced back to the different energy dependence of $a(k)$ for the 
$v_4$ and $v_6$ potential.

\section{$N$ particles in a trap}
\label{sectionIV}
To solve the time-independent Schr\"odinger equation for
more than $N=2$ trapped particles, we resort to the 
variational Monte Carlo (VMC) and diffusion Monte Carlo (DMC)
techniques~\cite{hammond}. 

In the VMC method, the variational many-body wave function $\psi_V$
is written in terms of a set of variational parameters 
$\mathbf{p}$,
which are optimized so as to minimize the 
variational energy $E_V$, 
$E_V= \langle \psi_V | H | \psi_V \rangle / \langle \psi_V | \psi_V \rangle$.
The energy expectation value $E_V$ is calculated for a given 
$\mathbf{p}$
using Metropolis sampling.  
Motivated by the structure of the Hamiltonian $H$, Eq.~(\ref{hb}), 
we write $\psi_V$ as a product
of one-body terms $\varphi$ and two-body Jastrow terms 
$F$~\cite{jastrow,dubois,blume01},
\begin{equation}
\psi_V(\mathbf{r}_1,\ldots,\mathbf{r}_N)=\prod_{i=1}^N \varphi(r_i)
\prod_{i<j}^N F(r_{ij}),
\end{equation}
where $\varphi(r)=\exp(-p_1 r^{p_2})$.
The functional form of the
two-body Jastrow factor $F$ is motivated by the functional
form of the 
interaction potential $v$ and by the fact
that we are interested in describing
the energetically lowest-lying gas-like state of the many-body Hamiltonian.
We use
\begin{equation}
\label{Fvn}
F(r) = \left\{
\begin{array}{c l}
  (1-b/r)(1+p_3/r^{p_4})  &  \mbox{ for } \quad  r>b \,  \\
  0     &  \mbox{ for } \quad  r \le b 
\end{array}
\right.
\end{equation}
for $v=v_n$, and 
\begin{equation}
\label{Fvps}
F(r) = \left\{
\begin{array}{c l}
  0  &  \mbox{ for } \quad  r \le c \,  \\
  \sin(kr+d)  &  \mbox{ for } \quad  c < r \le p_5 \,  \\
  e_1 + e_2 \exp(-p_6 r)  &  \mbox{ for } \quad  r > p_5 
\end{array}
\right.
\end{equation}
for $v=v_{ps}$.
The parameters $b$, $c$, $d$ and $k$ are chosen so that 
$\psi_V$ obeys the boundary conditions implied by the many-body
Hamiltonian $H$ and so that $\psi_V$ has the desired
symmetry (see below), while the parameters $e_1$ and $e_2$ are
determined by requiring that $F$ and its derivative are continuous
at $r=p_5$. For each interaction potential $v$, we optimize 
the variational parameters 
$\mathbf{p}$,
i.e., $p_1$ through $p_4$ for $v=v_n$,
and $p_1$, $p_2$, $p_5$ and $p_6$ for $v=v_{ps}$. 

To go beyond the variational calculations,
we apply two different variants of the DMC algorithm
both of which use the
optimized wave function $\psi_V$ as a guiding function~\cite{hammond}.
When the many-body Hamiltonian does not support any states
with negative energy,
the lowest lying gas-like state coincides with the true ground state
of the system. In this case, $\psi_V$ is nodeless and the DMC algorithm
with importance sampling results in the exact many-body energy.
When the many-body Hamiltonian supports negative energy states, i.e.,
molecular-like bound states, the energetically lowest lying 
gas-like state possesses nodes, which are imposed in the DMC method with
importance sampling through the
variational wave function $\psi_V$. This DMC variant, 
referred to as fixed-node DMC method \cite{hammond,reynolds},
determines the lowest energy of a state that has 
the same symmetry as $\psi_V$. Importantly, the FN-DMC energy provides
an upper bound to the exact eigenenergy of the excited gas-like state of
the many-body system \cite{reynolds}.

\begin{figure}[hbtp]
\vspace{0.25in}
\includegraphics[scale=0.26]{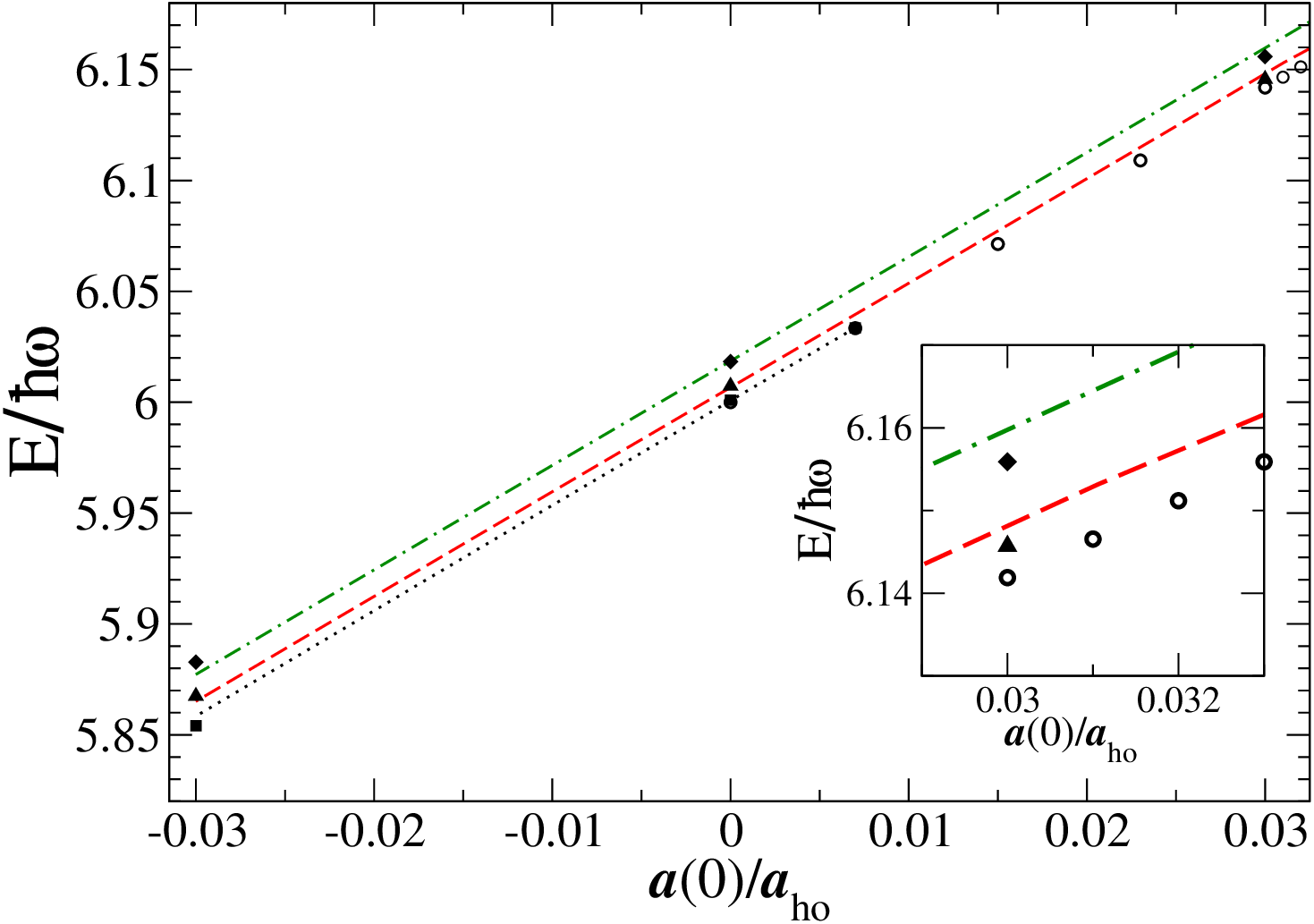}
\caption{\label{cap4}
(Color online)
Solid squares, triangles
and diamonds show the total
energy $E_{v_4}(N)$ calculated by the DMC method 
for four trapped particles interacting through the $v_4$ potential
with $r_c=0.007a_{ho}$ 
as a function
of $a(0)/a_{ho}$.  
The dotted, dashed, and dash-dotted lines
show the perturbative energies, Eq.~(\ref{pert}), 
for two-body potentials that support $0$, $1$
and $2$ bound states.
Open circles show the total energy $E_{a(0)}(N)$ 
calculated by the DMC method
for four particles interacting through
the zero-range pseudo-potential $v_{ps}$ with
$g=a(0)$.  
For all cases, the error bars of the energies
are smaller than the symbol size.
The inset shows 
an enlargement of the region around $a(0)= 0.03 a_{ho}$.
}
\end{figure}

The solid squares, triangles
and diamonds in Fig.~\ref{cap4}
show the total energy $E_{v_4}(N)$ calculated by the DMC method
for $N=4$ trapped particles interacting
through the $v_4$ potential with $r_c=0.007a_{ho}$
as a function of the zero-energy scattering length  $a(0)$
for a varying number of two-body $s$-wave bound states.
The solid 
squares
show energies for two-body potentials that support no $s$-wave
bound state. 
For the two-body potentials considered, the corresponding
four-body system supports no state with negative energy, 
and the hardcore boundary condition implied by $v_4$ is met
by setting the parameter $b$ in Eq.~(\ref{Fvn}) equal to $r_c$.
The energies for those $v_4$ potentials that support one and two $s$-wave
bound states are shown
by solid triangles and diamonds, respectively.
In these cases, the four-particle system supports negative energy states,
and the parameter $b$ 
is chosen to coincide with the $r$ value at which 
the free-space zero-energy 
two-body scattering solution has its first and second
node, respectively.
This construction of the 
many-body nodal surface assumes that
at most two particles scatter at any given time,
and that the nodal line of the two-body scattering solution is not
modified by the presence of the other 
atoms~\cite{blumeJPB,gao06}. 
This ``binary approximation'' is expected to 
be quite accurate in the low-density regime considered
throughout this paper.
Figure~\ref{cap4} shows that,
for a given zero-energy scattering length $a(0)$, the energy $E_{v_4}$ of 
the lowest-lying gas-like state 
increases as the number of two-body $s$-wave bound states,
or equivalently $\beta_4/a_{ho}$, increases, 
similar to the behavior found in Sec.~\ref{sectionII} for the
two particle case [see Fig.~\ref{cap1}(c)].
 
For comparison,
we consider the energy 
of $N$ particles interacting 
through an energy-dependent zero-range 
pseudopotential,
within first order perturbation theory,
\begin{equation}
\label{pert}
\frac{E(N)}{\hbar\omega}=\frac{3}{2}N+\frac{N(N-1)}{2}\sqrt{\frac{2}{\pi}}\;
\frac{a(k)}{a_{ho}}.
\end{equation}
For weakly interacting systems, i.e., 
for small scattering lengths, $a(k)$ can be
approximated by Eq.~(\ref{aEr4x}) with $k$ corresponding to the
trap energy scale of $3/2\hbar\omega$. 
The perturbative results for $N=4$ particles are 
shown in Fig.~\ref{cap4} 
by dotted, dashed
and dash-dotted lines for the cases when the two-body
potential supports zero, one
and two bound states.  
The agreement between the perturbative and 
DMC energies is reasonably 
good over the range of scattering lengths considered. In
particular, the perturbative expression with the energy-dependent
scattering length predicts the
up-shift of the energies with increasing number of two-body bound states
for a fixed $a(0)$ correctly but does not fully capture the change of slope
of the energy with increasing $|a(0)|$.

Next we consider the DMC results for $N$ trapped atoms interacting
through the pseudo-potential $v_{ps}$ with $g=a(0)$.
The potential $v_{ps}$ possesses one bound state for $g>0$ and
no bound state for $g<0$.  For positive scattering  
lengths, we use the nodal surface of the free-space two-body 
scattering solution for $v_{ps}$ to determine the
parameters of the Jastrow factor $F$, Eq.~(\ref{Fvps}), 
i.e., we use $c=a(0)$ and $d=\delta(k)$ with a very small $k$ value.
In addition to positive $a(0)$, we consider negative $a(0)$.
Using $\psi_V$ with $c$, $d$ and $k$ chosen so
that the boundary condition implied by the 
zero-range pseudo-potential is satisfied
whenever one of the interparticle distances $r_{ij}$ is zero, 
numerical instabilities associated with large negative DMC
energies arise.  These instabilities are 
most likely
associated with the Thomas collapse \cite{thomas,fedorov},
which is known to occur for systems with three or more particles
interacting through $v_{ps}$ with $g<0$.

The open circles
in Fig.~\ref{cap4} show the 
total DMC energy $E_{a(0)}(4)$
as a function of the zero-energy scattering length $a(0)$
for $a(0) \ge 0$.
We find that $E_{a(0)}(4)$ is smaller than or equal to
$E_{v_4}(4)$ for all $a(0)$. 
To quantify to which extent the Hamiltonian with the
shape-independent potential reproduces the properties
of the Hamiltonian with the shape-dependent potential, 
we consider the energy difference
$\Delta E(N)$,
$\Delta E(N) = E_{v_4}(N) - E_{a(0)}(N)$.
For $N=4$ and $10$, we find that $\Delta E$ scales---as might 
be expected for a weakly-interacting Bose gas---with the number
of pairs, i.e.,
$\Delta E(N) / \hbar \omega
\approx N_{pair} \Delta\varepsilon_{v_4}$, 
where $\Delta\varepsilon_{v_4}$ denotes the energy difference
introduced in Sec.~III and $N_{pair}$ the number of pairs, 
$N_{pair}=N(N-1)/2$.
For $N=4$ and $a(0)=0.03 a_{ho}$, e.g., 
we find $E_{v_4}= 6.1457(2) \hbar \omega$ 
(triangles in the inset of Fig.~4)
for the $v_4$ potential that
supports one two-body $s$-wave bound state
and
$E_{a(0)} = 6.14189(3) \hbar \omega$ 
(open circles in the inset of Fig.~4)
for the zero-range potential with $g=a(0)$,
and thus $\Delta E=0.0038(2) \hbar \omega$. For comparison,
the corresponding quantity
$N_{pair} \Delta\varepsilon$  equals $0.0040\hbar \omega$.
In addition to the $v_4$ potential, we 
consider the $v_6$ potential. 
In this case, the energy difference 
$\Delta E$ for comparatively small $a(0)$ is of the same order or larger than
the statistical uncertainties of our DMC energies and, although
expected to be valid, we
cannot explicitly
confirm the scaling of $\Delta E$ with $N_{pair}$ for the $v_6$
potential.

To further understand the 
implications of the energy-dependent scattering length $a(E)$ for $N>2$,
we determine the $a(E)$ that, if used to parametrize the interaction strength
of the zero-range potential in the many-body Hamiltonian, 
reproduces the energy
$E_{v_4}$.  For example, to reproduce the four-particle
energy $E_{v_4}=6.1457(2)$ for the $v_4$ potential with $a(0)=0.03 a_{ho}$
that supports 
one bound state,
the strength of the pseudopotential 
has to be $a(E)=0.03082(4)$.  
For the $v_4$ potential, this $a(E)$
corresponds to a two-body scattering energy of 
$1.53(16)\hbar\omega$.
Thus, the relevant scattering energy for two-body collisions
occuring in the weakly-interacting many-body 
is, not unexpectedly, approximately given by
the trap energy scale of $~3/2\hbar\omega$.

\section{Conclusion}
\label{sectionV}
This paper studies
trapped bosons interacting through attractive power-law
potentials with $r^{-4}$ and $r^{-6}$ tails.  
For two particles, the energy-dependent pseudo-potential
accurately reproduces the 
energies for both shape-dependent potentials.  
Further, we find that the deviations between the energies obtained for the
energy-independent pseudo-potential and the shape-dependent
potential scale
for small $a(0)$  
as $(\beta_6/a_{ho})^3$ and
$(\beta_4/a_{ho})^2$ for the potentials with $r^{-6}$ and $r^{-4}$ tails,
respectively.  Finally, we use Monte Carlo methods to extend the treatment to more than
two trapped particles.  
Again, we find that the energy for the shape-dependent power-law
potential can be reproduced accurately by the energy-dependent 
pseudo-potential if the energy-scale entering the pseudo-potential
is chosen properly. In general, this leads to a self-consistent
many-body framework that considers only binary interactions
but includes many-body correlations.

Even if the $r^{-4}$ results are not directly applicable 
to present-day experiments
(combined atom-ion systems have not yet been trapped),
our comparative study of the
energetics for $r^{-4}$ and $r^{-6}$ potentials 
provides insights into weakly interacting systems in general 
and van der Waals $r^{-6}$ interactions in particular.
Our calculations suggest that three-body terms \cite{wu} are very small in 
the dilute limit considered throughout this work.
It seems feasible
that the description of systems with longer-ranged interactions 
at the mean-field level can be improved
by including the energy dependence
of the scattering length, similar to the frameworks outlined in
Refs.~\cite{gao03,collin}.

We gratefully acknowledge support by the NSF through Grant No. PHY-0555316.

\end{document}